\documentclass[
superscriptaddress,
nofootinbib,
twocolumn,
amsmath,amssymb,
aps,
pra,
notitlepage,
longbibliography
]{revtex4-1}

\usepackage{dcolumn}
\usepackage{bm}
\usepackage{epsfig}
\usepackage{graphicx}
\usepackage{latexsym}
\usepackage{amsfonts}
\usepackage{float} 
\usepackage{setspace}
\usepackage{graphicx}
\usepackage{amsmath}
\usepackage{mathrsfs}
\usepackage{verbatim}
\usepackage{color}
\usepackage{SIunits}
\usepackage{hyperref}
\usepackage{multirow}
\usepackage{bigstrut}
\usepackage{array}
\usepackage{tabularx}
\usepackage{physics}
\usepackage{ragged2e}

\newcommand{\be}{\begin{equation}}
	\newcommand{\ee}{\end{equation}}
\newcommand{\ba}{\begin{array}}
	\newcommand{\ea}{\end{array}}
\newcommand{\bqa}{\begin{eqnarray}}
	\newcommand{\eqa}{\end{eqnarray}}

\begin{document}
\title{Perspectives on epitaxial InGaP for quantum and nonlinear optics}

\author{Joshua Akin} 
\thanks{These authors contributed equally to this work.}
\affiliation{Holonyak Micro and Nanotechnology Laboratory and Department of Electrical and Computer Engineering, University of Illinois at Urbana-Champaign, Urbana, IL 61801 USA}
\affiliation{Illinois Quantum Information Science and Technology Center, University of Illinois at Urbana-Champaign, Urbana, IL 61801 USA}
\author{Yunlei Zhao} 
\thanks{These authors contributed equally to this work.}
\affiliation{Holonyak Micro and Nanotechnology Laboratory and Department of Electrical and Computer Engineering, University of Illinois at Urbana-Champaign, Urbana, IL 61801 USA}
\affiliation{Illinois Quantum Information Science and Technology Center, University of Illinois at Urbana-Champaign, Urbana, IL 61801 USA}
\author{A. K. M. Naziul Haque} 
\affiliation{Holonyak Micro and Nanotechnology Laboratory and Department of Electrical and Computer Engineering, University of Illinois at Urbana-Champaign, Urbana, IL 61801 USA}
\affiliation{Illinois Quantum Information Science and Technology Center, University of Illinois at Urbana-Champaign, Urbana, IL 61801 USA}
\author{Kejie Fang} 
\email{kfang3@illinois.edu}
\affiliation{Holonyak Micro and Nanotechnology Laboratory and Department of Electrical and Computer Engineering, University of Illinois at Urbana-Champaign, Urbana, IL 61801 USA}
\affiliation{Illinois Quantum Information Science and Technology Center, University of Illinois at Urbana-Champaign, Urbana, IL 61801 USA}

\begin{abstract} 

Nonlinear optical materials are essential for the development of both nonlinear and quantum optics and have advanced recently from bulk crystals to integrated material platforms. In this Perspective, we provide an overview of the emerging InGaP $\chi^{(2)}$ nonlinear integrated photonics platform and its experimental achievements. With its exceptional $\chi^{(2)}$ nonlinearity and low optical losses, the epitaxial InGaP platform significantly enhances a wide range of second-order nonlinear optical effects, from second-harmonic generation to entangled photon pair sources, achieving efficiencies several orders of magnitude beyond the current state of the art. Moreover, the InGaP platform enables quantum nonlinear optics at the few- and single-photon levels via passive nonlinearities, which has broad implications for quantum information processing and quantum networking. We also examine the current limitations of the InGaP platform and propose potential solutions to fully unlock its capabilities.

\end{abstract}
	
\maketitle

\section{INTRODUCTION}
 
Nonlinear optical materials play a pivotal role in advancing both nonlinear optics and quantum optics. In addition to their classical uses, such as in optical parametric oscillators \cite{giordmaine1965tunable} and high-harmonic generation \cite{ghimire2011observation, you2017anisotropic}, nonlinear crystals have long been employed to generate quantum light, including entangled photon pairs \cite{kwiat1995new} 
and squeezed light \cite{schnabel2017squeezed}. 
These quantum light sources are foundational to the experimental progress in quantum communications \cite{gisin2007quantum}, quantum sensing \cite{lawrie2019quantum}, and quantum computing \cite{o2007optical}.

\begin{figure}[!htb]
	\begin{center}
		\includegraphics[width=0.8\columnwidth]{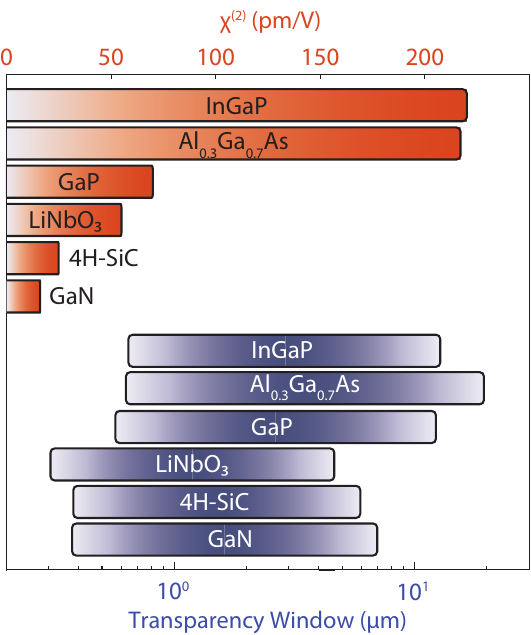}
		\caption{$\chi^{(2)}$ susceptibility and transparency window of leading thin-film second-order nonlinear materials. }
		\label{fig:TW}
	\end{center}
\end{figure}

In recent years, nonlinear optical materials have progressed from bulk crystals to integrated material platforms, driven by advances in material growth and fabrication techniques. These developments have facilitated the use of materials with higher nonlinear susceptibilities and the creation of light-confining nanophotonic structures, resulting in a substantial increase in nonlinear optical efficiencies \cite{dutt2024nonlinear}. Fig. \ref{fig:TW} summarizes several leading second-order ($\chi^{(2)}$) nonlinear thin-film materials, including their second-order susceptibility and transparency window. Among them, III-V materials, including Al$_x$Ga$_{1-x}$As and In$_{0.5}$Ga$_{0.5}$P (InGaP), exhibit the highest $\chi^{(2)}$ susceptibilities and a broad transparency window.  They can be epitaxially grown on the GaAs substrate, leading to versatile III-V integrated photonics platforms. In the case of Al$_x$Ga$_{1-x}$As, increasing the aluminum composition expands the bandgap, but this also results in a marked decrease in the second-order susceptibility \cite{ohashi1993determination}. Moreover, arsenic-based III-V materials suffer from strong optical absorption at wavelengths below 800 nm \cite{michael2007wavelength, placke2024telecom}, due to the antibonding As-As surface state that lies below the bandgap \cite{lin2012passivation}. These limitations have made Al$_x$Ga$_{1-x}$As a less suitable choice for second-order nonlinear optics involving light in the telecommunication band and the corresponding second harmonics.

\begin{figure*}[!htb]
	\begin{center}
		\includegraphics[width=2\columnwidth]{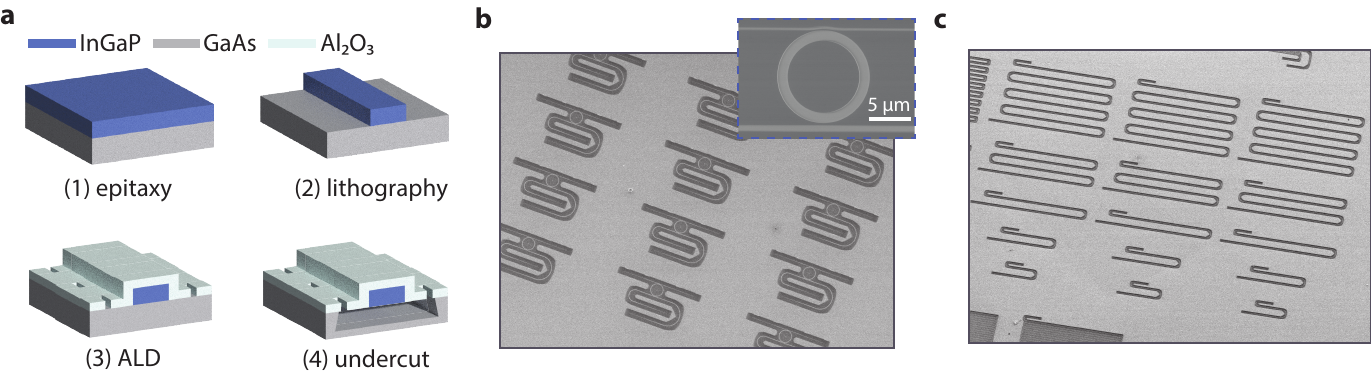}
		\caption{\textbf{a}. Illustration of the fabrication process of the suspended InGaP photonic integrated circuit. \textbf{b}. SEM images of waveguide-coupled microring resonators. \textbf{c}. SEM image of waveguide devices.}
		\label{fig:fab}
	\end{center}
\end{figure*}

In this Perspective, we provide an overview of the emerging InGaP $\chi^{(2)}$ nonlinear integrated photonics platform and discuss its unique advantages for nonlinear and quantum optics involving the telecommunication band. In Sec. \ref{Sec:II} we discuss the material properties of InGaP. Section \ref{Sec:III} discusses properties of InGaP nonlinear photonic devices, including microring resonators and waveguides. Sections \ref{Sec:IV} and \ref{Sec:V} discuss nonlinear and quantum optical effects of the cavity and waveguide, respectively. Finally, we briefly examine the current limitations of the InGaP platform and propose potential solutions.

\section{INGAP AS A PRIME NONLINEAR MATERIAL}\label{Sec:II}

InGaP is a III-V semiconductor material that is lattice-matched to GaAs. As a result,  thin-film single-crystalline InGaP can be epitaxially grown on the GaAs substrate via metal-organic chemical vapor deposition (MOCVD) or molecular beam epitaxy (MBE). Due to its high electron mobility, direct bandgap and thermal stability, InGaP has been used for applications including solar cells \cite{takamoto2005ingap}, LEDs \cite{svensson2008monolithic}, photon detectors \cite{jiang2004high},  heterojunction bipolar transistors \cite{pan1998high} and high-electron-mobility transistors \cite{zheng2000growth}. In addition, thin-film InGaP has also been studied for nonlinear optics by exploiting its large Kerr nonlinearity \cite{eckhouse2010highly, colman2010temporal}, leading to the demonstration of frequency combs \cite{dave2015dispersive}, optical parametric oscillators \cite{marty2021photonic}, and entangled photon generation via four-wave mixing \cite{chopin2023ultra}.

Remarkably, InGaP also possesses a large second-order susceptibility ($\chi^{(2)}_{xyz}=220$ pm/V) \cite{ueno1997second} and a bandgap of 1.92 eV corresponding to a cutoff wavelength of 645 nm. This marks InGaP as another thin-film material, besides Al$_x$Ga$_{1-x}$As ($x\approx 0.3$), that has the highest second-order susceptibility as well as a cutoff wavelength less than the second harmonics of the telecommunication C band light. However, what makes InGaP unique for second-order nonlinear optics involving the telecommunication C band light is the fact that its antibonding anion state lies well above the bandgap \cite{lin2012passivation}, which avoids anion state-induced absorption below the bandgap. In contrast, the antibonding As-As surface state of Al$_x$Ga$_{1-x}$As lies below its bandgap \cite{lin2012passivation}, resulting in strong optical absorption at wavelengths shorter than 800 nm \cite{michael2007wavelength, placke2024telecom}. In addition, the transparency window of InGaP extends to $\sim$11 $\mu$m \cite{wilson2020integrated}, which will enable applications in the mid-infrared wavelength band, including optical parametric oscillators \cite{marandi2012coherence,vainio2016mid}, detectors \cite{razeghi2014advances}, and quantum cascade lasers \cite{yao2012mid}.

\section{INGAP NONLINEAR PHOTONIC INTEGRATED CIRCUITS}\label{Sec:III}

Because InGaP has a similar refractive index as the GaAs substrate from which it is grown, in order to confine light in the InGaP thin-film photonic structures, it needs to be separated from the GaAs substrate. This can be achieved by transferring the InGaP thin film to a low-index substrate using various wafer bonding methods, including adhesive bonding \cite{dave2015nonlinear} and, more recently, low-temperature plasma-activated bonding \cite{thiel2024wafer}. However, wafer bonding or flip-chip bonding can be challenging, requiring special tools and complicated processing. Instead, we developed a transfer-free method that makes suspended InGaP photonic integrated circuits in the InGaP-on-GaAs platform. The fabrication process is illustrated in Fig. \ref{fig:fab}a. Starting from the  InGaP-on-GaAs epitaxial stack, the device pattern is defined using electron beam lithography and transferred to the InGaP thin film via inductively coupled plasma reactive-ion etch (ICP-RIE) using a mixture of Cl$_2$/CH$_4$/Ar gas. Then a thin layer of Al$_2$O$_3$ is deposited on the sample via atomic layer deposition (ALD) followed by patterning the releasing holes in the Al$_2$O$_3$ layer. Finally, the InGaP device is released from the GaAs substrate using selective wet etching. More details of the fabrication process are provided in Refs. \cite{zhao2022ingap, akin2024ingap}.  Figs. \ref{fig:fab}b and c show scanning electron microscope (SEM) images of the fabricated InGaP photonic integrated circuits, including waveguide-coupled microring resonators and meander waveguides. 

\begin{figure*}[!htb]
	\begin{center}
		\includegraphics[width=2\columnwidth]{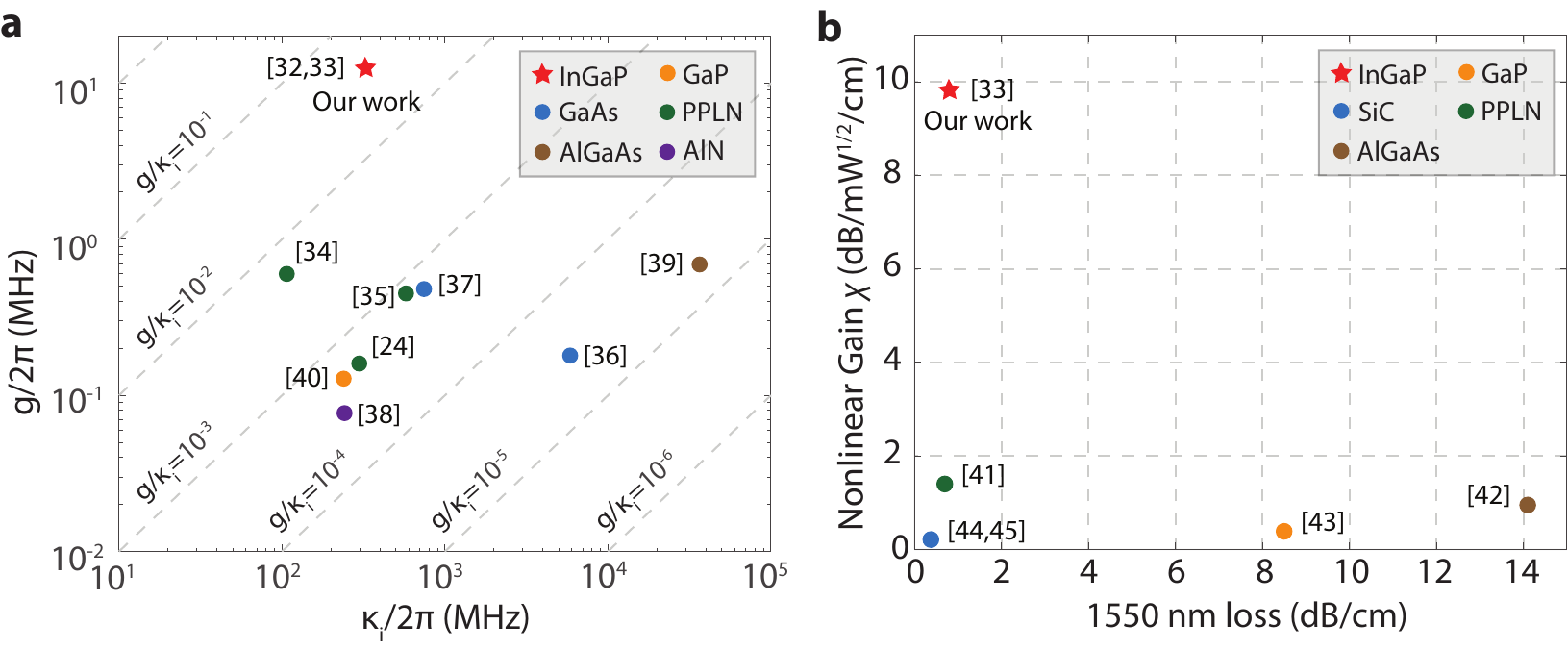}
		\caption{\textbf{a}. Nonlinear mode coupling coefficient ($g$) and intrinsic photon loss rate of the fundamental mode ($\kappa_i$) of microring resonators made in second-order nonlinear photonics platforms, including InGaP: \cite{zhao2022ingap, akin2024ingap}, PPLN: \cite{lu2020toward,ma2020ultrabright,ueno1997second}, GaAs: \cite{kuo2014second,chang2019strong}, AlN: \cite{bruch201817}, AlGaAs: \cite{mariani2014second}, GaP: \cite{logan2018400}. \textbf{b}. Nonlinear gain $\chi$ and 1550 nm wavelength band loss of nanophotonic waveguides made in second-order nonlinear photonics platforms. InGaP: \cite{akin2024ingap}, PPLN: \cite{wang2018ultrahigh}, AlGaAs: \cite{may2019second}, GaP: \cite{pantzas2022continuous}, SiC: \cite{zheng2022efficient,yi2022silicon}. }
		\label{fig:SHG and loss}
	\end{center}
\end{figure*}

\subsection{Microring resonators}

In order to exploit the second-order susceptibility $\chi_{xyz}^{(2)}$ of InGaP for second-order nonlinear optics, transverse-electric modes (TE$_{00}$) and transverse-magnetic modes (TM$_{00}$) of the microring resonator are used for the fundamental- and second-harmonics, respectively. The fundamental-harmonic (FH) and second-harmonic (SH) modes, $a$ and $b$, satisfy the frequency- and phase-matching conditions \cite{zhao2022ingap}: $\omega_b=2\omega_a$ and $m_b=2m_a\pm 2$, where $m_k$ is the azimuthal number of mode $k$. This can be achieved by designing the width of the microring for a given InGaP thickness. For phase matching between the telecommunication band FH mode and the corresponding SH mode, the thickness of InGaP should be around 110 nm \cite{zhao2022ingap}.

The quality factor of the InGaP microring resonator depends on the ALD material used to suspend the microring resonator  \cite{akin2024ingap}. The intrinsic quality factor ($Q_{ai}$) of 1550 nm band TE$_{00}$ resonances of the fabricated microring with 5 $\mu$m radius and 1 $\mu$m width is $4-6\times 10^5$ for ALD Al$_2$O$_3$, which is approximately $2-3\times$ higher than that of microring resonators with ALD SiO$_2$. This is likely due to the passivation effect of Al$_2$O$_3$ \cite{guha2017surface}. The intrinsic quality factor of 775 nm band TM$_{00}$ resonances of the same microring is $1-2\times 10^5$, and is nearly independent of the ALD material. 

The nonlinear interaction between the FH and SH modes can be described by the Hamiltonian, $\hat H=g(\hat a^2\hat b^\dagger+\hat a^{\dagger 2}\hat b)$, where $\hat a(\hat b)$ and $\hat a^\dagger(\hat b^\dagger)$ are the annihilation and creation  operators for mode $a(b)$. The nonlinear coupling $g$ is inferred from the second-harmonic generation (SHG) efficiency of the microring.  For the phase-matched 5 $\mu$m-radius microring, we find $g/2\pi\approx 11$ MHz \cite{zhao2022ingap}, which is consistent with the simulation. The nonlinear coupling scales with the microring radius approximately as $g\propto\frac{1}{\sqrt{R}}$.

A nonlinearity-to-loss ratio, $g/\kappa_{ai}$, where $\kappa_{ai}\equiv\omega_a/Q_{ai}$, can be defined to characterize the nonlinear cavity. Almost all second-order nonlinear optical effects of the cavity are determined by this ratio. Table \ref{tab:cavityprocess} summarizes the relation between the key metrics of several nonlinear optical processes and the nonlinearity-to-loss ratio. The 5 $\mu$m-radius InGaP microring with $Q_{ai}=6\times 10^5$ achieves a high $g/\kappa_{ai}=3.5\%$. In Fig. \ref{fig:SHG and loss}a, we compare the nonlinearity-to-loss ratio of the state-of-the-art InGaP microring with microring resonators made in other $\chi^{(2)}$ nonlinear photonics platforms.

{\renewcommand{\arraystretch}{1.3}
\begin{table}[h!]
\centering
\caption{\textbf{Cavity $\chi^{(2)}$ process }}
\vspace{5pt}
\begin{tabularx}{0.49\textwidth} {m{2.9cm}  m{3.4cm}  m{2.5cm} }
    \hline
    \hline
       $\chi^{(2)}$ process& Metric &  Scaling  \\
       \hline
       SHG & Efficiency &  $g^2/\kappa_{ai}^2\kappa_{bi}$ \cite{guo2016second}   \\
       SPDC &  Pair generation rate & $g^2/\kappa_{ai}\kappa_{bi}$ \cite{guo2017parametric} \\ 
       OPO   & Pump threshold  &  $\kappa_{ai}^2\kappa_{bi}/g^2$ \cite{walls2007quantum}   \\
       Squeezed light & Pump power  &  $\kappa_{ai}^2\kappa_{bi}/g^2$ \cite{walls2007quantum} \\
    \hline
    \hline
    \end{tabularx}
    
{\justifying \footnotesize  SPDC: spontaneous parametric down-conversion. OPO: optical parametric oscillation. Pump power for squeezed light is assumed for optimal quadrature noise squeezing (i.e., $\kappa_{a,i}/\kappa_a$) on resonance. Critical coupling ($\kappa_i=\kappa/2$) is assumed. \par }
    \label{tab:cavityprocess}
\end{table}

{\renewcommand{\arraystretch}{1.5}
\begin{table*}[!htb]
	\centering
	\caption{\textbf{Waveguide $\chi^{(2)}$ process}}
	\begin{tabular*}{\linewidth}{@{\extracolsep{\fill}} ccc }
		\hline\hline
		$\chi^{(2)}$ process & Metric & Relation \bigstrut\\
		\hline
		\multirow{2}[4]{*}{SPDC} & Pair generation efficiency & $\frac{\hbar\omega_pL^{3/2}}{12\sqrt{2\pi\left|\mathrm{GVD}\left(\frac{\omega_p}{2}\right)\right|}}\chi^2$ \cite{kumar2018parametric} \bigstrut\\
		\cline{2-3}          & Bandwidth & $\frac{\alpha}{\sqrt{\left|\mathrm{GVD}\left(\frac{\omega_p}{2}\right)\right|L}}$ \cite{kumar2018parametric} \bigstrut\\
		\hline
		\multirow{2}[4]{*}{OPA} & Gain  & $e^{\chi \sqrt{P_P} L}$ (maximum gain of PSA);  $\frac{1}{4}e^{\chi \sqrt{P_P} L}$ (PIA) \bigstrut\\
		\cline{2-3}          & Bandwidth & $\frac{\alpha}{\sqrt{\left|\mathrm{GVD}\left(\frac{\omega_p}{2}\right)\right|L}}$ \bigstrut\\
		\hline
		Squeezed light & Squeeze ratio & $\epsilon \ e^{-\chi \sqrt{P_P} L} + 1 - \epsilon$ \cite{stokowski2023integrated} \bigstrut\\
		\hline
		Wavelength conversion & Conversion efficiency & $\sin^2(\chi\sqrt{P_p}L/\sqrt{2})$ \bigstrut\\
		\hline\hline
	\end{tabular*}%
\label{tab:waveguideprocess}

{\justifying \footnotesize GVD: group velocity dispersion. $\alpha=\frac{1}{\pi}\sqrt{2\mathrm{sinc}^{-1}\frac{1}{\sqrt{2}}}$. OPA: optical parametric amplification. PSA: phase-sensitive amplifier. PIA: phase-insensitive amplifier. $\epsilon=e^{-\gamma L}$: aggregated waveguide loss. The waveguide loss is only considered for the squeezed light.   \par}
\end{table*}
\vspace{-12pt}
\vspace{6pt}

\subsection{Waveguides}

Broadband second-order nonlinear optical effects can be realized in waveguides as opposed to microring resonators. Similar to the microring resonator, TE$_{00}$ and TM$_{00}$ modes of the waveguide are used for the fundamental- and second-harmonics, respectively. By designing the waveguide width for a given InGaP thickness, the two modes can satisfy the frequency- and phase-matching conditions: $\omega_b=2\omega_a$ and $k_b=2k_a$, which lead to $n_b=n_a$, where $n_k$ is the effective mode index of mode $k$. 

The phase-matched nonlinear waveguide can be characterized by its loss and nonlinear efficiency. The loss of the TE$_{00}$ mode of the InGaP waveguide in the 1550 nm band is measured to be $\gamma=0.8\pm0.4$ dB/cm \cite{akin2024ingap}, which is consistent with the intrinsic quality factor of the microring resonator (corresponding to 0.4 dB/cm). The difference might be due to the uncertainty of the fiber-optic coupling efficiency when measuring the waveguide transmission. The normalized SHG efficiency of the phase-matched InGaP waveguide is measured to be 128,000 $\%$/W/cm$^2$ \cite{akin2024ingap}, which is consistent with the simulation value of 130,000 $\%$/W/cm$^2$. This is significantly higher than other demonstrations of InGaP $\chi^{(2)}$ nonlinear waveguides \cite{poulvellarie2021efficient,peralta2022low}, which are limited by considerable optical losses and imperfect phase matching.  The SHG efficiency of our InGaP nanophotonic waveguides is nearly two orders of magnitude higher than the thin-film periodically-poled LiNbO$_3$ (PPLN) waveguides in the telecommunication band \cite{wang2018ultrahigh,chen2024adapted}. Such an enhancement in nonlinear efficiency can be understood via a back-of-the-envelope calculation. The normalized SHG efficiency of the InGaP waveguide can be calculated using \cite{akin2024ingap}
\begin{equation}
	\begin{gathered}
		\eta_{\mathrm{SHG}} = \frac{\omega_a^2}{2n_a^2n_b\epsilon_0 c^3} \left( \frac{\int \,d\textbf{r} \chi_{xyz}^{(2)} \sum\limits_{i\neq j\neq k}E_{bi}^*E_{aj}E_{ak}}{\int \,d\textbf{r}\abs{\textbf{E}_a}^2\sqrt{\int \,d\textbf{r}\abs{\textbf{E}_b}^2}} \right) ^2,
	\end{gathered}
	\label{eq:PM}
\end{equation}
where the normalization integrals use electric field components perpendicular to the wavevector of the waveguide mode. Based on Eq. \ref{eq:PM}, the ratio between the nonlinear efficiency of InGaP and PPLN waveguides is estimated to be: $(4\times2\times\frac{\pi}{2})^2\approx160$, where $4\times$ is from $\chi^{(2)}_{xyz}$ of InGaP versus $\chi^{(2)}_{zzz}$ of LiNbO$_3$, $2\times$ is from the swap of indices $x$ and $y$ in $\chi^{(2)}_{xyz}$ for InGaP, and $1/\frac{2}{\pi}\times$ is due to the periodic poling of LiNbO$_3$. This estimation does not consider the difference in the mode overlap integral, which is generally an $O(1)$ factor.

Using the normalized SHG efficiency, we can introduce a nonlinear gain defined as 
\be\label{nlgain}
\chi\equiv 2\sqrt{\eta_{\mathrm{SHG}}}.
\ee
The nonlinear gain is convenient to characterize second-order nonlinear optical processes in the waveguide, including optical parametric amplification and squeezed light generation. Table \ref{tab:waveguideprocess} summarizes the key metrics of several second-order nonlinear optical processes of the waveguide as functions of the nonlinear gain $\chi$ and waveguide loss.  Fig. \ref{fig:SHG and loss}b compares the nonlinear gain and waveguide loss of the InGaP platform with other $\chi^{(2)}$ nonlinear photonics platforms. In this plot, the numerical value of the nonlinear gain is calculated via $\chi\ [\mathrm{dB}/\sqrt{\mathrm{mW}}/\mathrm{cm}]=10\mathrm{log}(e^{2\sqrt{\eta_{\mathrm{SHG}}}})$. The InGaP waveguide achieves $\chi=9.8$ dB/$\sqrt{\mathrm{mW}}$/cm, which is significantly greater than other platforms.   \\

\section{QUANTUM OPTICS: CAVITY}\label{Sec:IV}

The nonlinearity-to-loss ratio of a few percent achieved in InGaP microring resonators enables quantum optical effects using passive bulk $\chi^{(2)}$  nonlinearity at the few- and single-photon levels. Creating quantum correlations between initially uncorrelated photons, such as photon antibunching and photon blockade, are fundamental quantum optical effects and useful quantum resources.  This is usually achieved using quantum emitters \cite{birnbaum2005photon} or quantum interference induced by ancillary pumps \cite{flayac2017unconventional}.  Recently it was theoretically shown that photon antibunching and photon blockade can be achieved via passive bulk nonlinearities that are not necessarily in the strong coupling regime \cite{wang2022few}. Consider two uncorrelated photons propagate through a waveguide-coupled $\chi^{(2)}$ nonlinear cavity. The wavefunction of the two photons after traversing the system consists of the nonlinear interaction-mediated amplitude $T$ and the linear transmission amplitude $t^2$, where $t$ is the single-photon transmission coefficient, leading to a normalized second-order correlation function \cite{wang2022few}:
\be\label{g2}
g^{(2)}(\tau)=\frac{|t_\omega^2+ T(\omega,\tau)|^2}{|t_\omega^2|^2},
\ee 
where $\omega$ denotes the frequency of the photons and $|T|\sim (g/\kappa)^2$. As a result, quantum correlations between photons, such as photon antibunching and photon blockade, can be achieved by controlling the single-photon transmission coefficient of the photonic circuit such that $t_\omega^2+ T(\omega,\tau)\approx 0$, even for $g/\kappa<1$. Initial experimental demonstration of such an effect has been realized using a waveguide-coupled InGaP microring resonator \cite{zhao2023quantum}, where photon antibunching (i.e., $g^{(2)}(0)<g^{(2)}(\tau)$ for some $\tau$) was observed. However, due to the free-standing InGaP device structure, thermal noises associated with the inherent mechanical vibrational modes of the device prevented the realization of photon blockade, i.e., $g^{(2)}(0)\approx 0$. We expect this issue to be solved by adopting the InGaP-on-insulator device structure, which is free of vibrational modes. 

The highly nonlinear InGaP cavity also enables nonlinear-optical entangling operations that facilitate quantum communications and quantum networking. Existing quantum networking protocols, such as quantum teleportation and entanglement swapping, rely on linear-optical Bell state measurements to herald the distribution and transfer of quantum information. However, the linear-optical Bell state measurement necessitates identical photons, making them prone to errors from multiphoton emissions \cite{kok2000postselected,pan2003experimental}. This dependency limits both the efficiency and fidelity of entanglement distribution. Additionally, any deviation in the input photons' identity can compromise the protocol's fidelity \cite{sangouard2011quantum,azuma2023quantum}. A solution to this issue is the nonlinear-optical Bell state measurement, which uses the sum-frequency generation (SFG) process to filter out multiphoton emissions. This approach also mitigates the need for quantum interference of identical photons; instead, the input photons only need to meet the phase-matching condition for sum-frequency generation. Deviations from this condition primarily impact efficiency rather than fidelity. Recently, as reported in \cite{akin2024faithful}, we demonstrated the nonlinear Bell state measurement by exploiting an InGaP nonlinear cavity with a single-photon SFG probability ($4\times10^{-5}$) over three orders of magnitude beyond prior nonlinear waveguides \cite{tanzilli2005photonic,guerreiro2013interaction,guerreiro2014nonlinear,fisher2021single}.
Using the nanophotonic nonlinear Bell state analyzer, we achieved faithful quantum teleportation involving time-bin encoded, spectrally-distinct photons with a fidelity $\geq94\%$ down to the single-photon level and validated the robustness of this scheme against multiphoton emission. Beyond quantum teleportation, a nonlinear Bell state analyzer with even moderate SFG efficiency can facilitate faithful heralded entanglement swapping \cite{sangouard2011faithful}---a crucial protocol for quantum repeaters---more efficiently than linear-optical protocols using ancillae \cite{wagenknecht2010experimental,barz2010heralded}.

\section{QUANTUM OPTICS: WAVEGUIDE}\label{Sec:V}

In this section, we analyze several nonlinear and quantum optical processes in waveguides and highlight the advantage of the InGaP platform. One application of the $\chi^{(2)}$ nonlinear waveguide is for entangled photon pair generation via SPDC. The relations of the pair generation efficiency ($P_{\mathrm{SPDC}}/P_p$) and the bandwidth of the SPDC photons of a phase-matched waveguide are given in Table \ref{tab:waveguideprocess}. Recently, we demonstrated an ultra-bright, broadband, time-energy entangled photon source utilizing the InGaP nanophotonic waveguide \cite{akin2024ingap}. A 1.6 mm long InGaP waveguide achieves a pair generation rate of 97 GHz/mW and a bandwidth of 14.4 THz (115 nm) centered at the telecommunication C band. This corresponds to a per-bandwidth pair generation efficiency of 6.7 GHz/mW/THz (840 MHz/mW/nm). The ultrahigh rate efficiency of the InGaP nanophotonic waveguide can be understood from the fact that the pair generation efficiency is proportional to the SHG efficiency (Table \ref{tab:waveguideprocess}). In Fig. \ref{fig4:ES}, we compare the InGaP nanophotonic waveguide SPDC source with other broadband SPDC sources, including bulk crystals and  waveguides. The InGaP waveguide SPDC source exhibits a leading pair generation efficiency with a large bandwidth, underscoring its potential for quantum information applications.

\begin{figure}[!htb]
	\begin{center}
		\includegraphics[width=0.9\columnwidth]{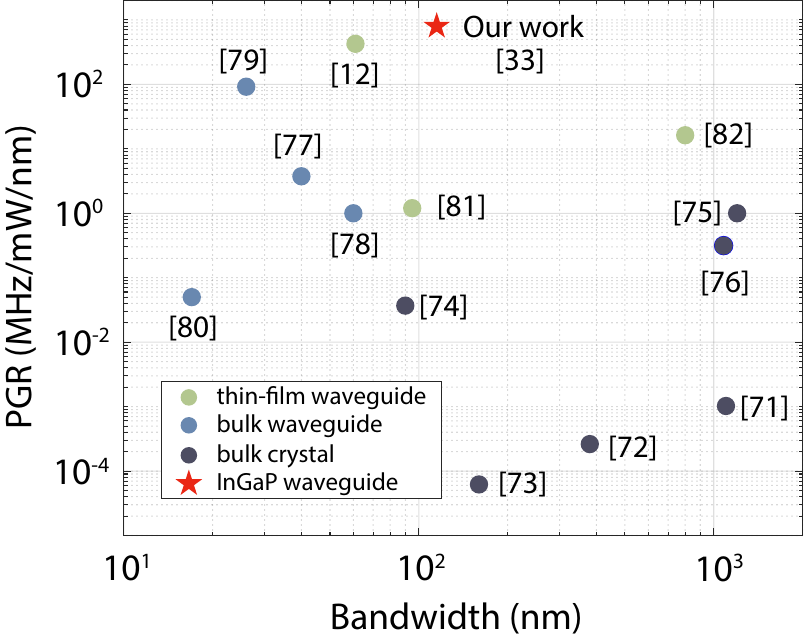}
		\caption{ Per-bandwidth pair generation rate (PGR) and bandwidth of recently demonstrated broadband SPDC sources. Bulk crystal: \cite{tanaka2012noncollinear,okano20150,okano2012generation,chekhova2018broadband,shaked2014observing,o2007observation}.  Bulk waveguide: \cite{tanzilli2002ppln,huang2022high,cao2023non,u2004efficient}. Thin-film waveguide: \cite{akin2024ingap,kang2016monolithic,javid2021ultrabroadband,placke2024telecom}. }
		\label{fig4:ES}
	\end{center}
\end{figure}

Another application of $\chi^{(2)}$ nonlinear waveguides is for traveling-wave optical parametric amplifiers (OPAs). In this case, a pump light of frequency $\omega_p$ and a signal light of frequency $\omega_s$, satisfying $\omega_p\approx2\omega_s$, are injected into the $\chi^{(2)}$ nonlinear waveguide. When the waveguide satisfies the phase-matching condition, the three-wave mixing process transfers energy from the pump to amplify the signal. For a waveguide that is phase-matched for the degenerate three-wave mixing process, the bandwidth of the traveling-wave OPA is identical to that of the SPDC photons (Table \ref{tab:waveguideprocess}), determined by the waveguide length and the group velocity dispersion of the waveguide. When the frequencies of the pump and signal satisfy $\omega_p=2\omega_s$, the OPA is phase-sensitive, i.e., the gain depends on the phase of the signal. In this case, the maximum gain is given by
\be
G_\mathrm{PSA}= e^{\chi\sqrt{P_p}L},
\ee
where $\chi$ is the nonlinear gain introduced in Eq. \ref{nlgain}, $P_p$ is the pump power, and $L$ is the length of the waveguide.  When the frequencies of the pump and signal do not satisfy the double relationship and the signal lies in the phase-matching bandwidth, the OPA is phase-insensitive and the constant gain is given by 
\be
G_\mathrm{PIA}=\cosh^2(\chi\sqrt{P_p}L/2)\approx\frac{1}{4}e^{\chi\sqrt{P_p}L}.
\ee
Essentially, the gain of a phase-insensitive OPA is 6 dB less than the maximum gain of a phase-sensitive OPA \cite{andrekson2020fiber}. From the measured nonlinear gain $\chi=9.8$ dB/$\sqrt{\mathrm{mW}}$/cm of the InGaP waveguide, we expect a maximum gain of 30 dB can be achieved in a 1 cm long InGaP waveguide with only 10 mW on-chip pump power. The InGaP OPA empowered by its extreme $\chi^{(2)}$ nonlinearity is expected to transcend the state-of-the-art integrated traveling-wave OPAs based on Kerr nonlinearity \cite{riemensberger2022photonic,kuznetsov2024ultra}, in terms of pump efficiency, device size, and noise figure.

Related to the OPA, $\chi^{(2)}$ nonlinear waveguides can be used for generating squeezed light. When the waveguide is pumped at frequency $\omega_p$, squeezed vacuum at the frequency of $\omega_p/2$ or two-mode squeezed light at frequencies $\omega_p/2\pm\Delta$ can be generated. The quantum fluctuation of one quadrature of the squeezed vacuum (or the sum of one quadrature of the two-mode squeezed light) is de-amplified while that of the orthogonal quadrature (or the sum of the orthogonal quadrature) is amplified. Waveguide loss needs to be considered for the squeezed light since it corresponds to the vacuum noise. The squeeze ratio $R$, defined as the reduction of the quadrature noise power, of the squeezed vacuum and the two-mode squeezed light generated in a phase-matched $\chi^{(2)}$ nonlinear waveguide is given by \cite{stokowski2023integrated}
\be\label{eq:squeeze}
R=\epsilon \ e^{-\chi \sqrt{P_P} L} + 1 - \epsilon,
\ee
where $\epsilon=e^{-\gamma L}$ is the aggregated waveguide loss. According to Eq. \ref{eq:squeeze}, for a given pump power, there exists an optimal waveguide length to achieve the maximum squeeze ratio, due to the trade-off between the parametric gain and waveguide loss. For example, using the measured nonlinear gain $\chi=9.8$ dB/$\sqrt{\mathrm{mW}}$/cm and waveguide loss of 0.8 dB/cm, for $P_p=10(100)$ mW, an optimal squeezing of 9.5(13.5) dB can be achieved with a 5(2) mm long InGaP waveguide. This predicted squeezing level is significantly higher than current integrated squeezed light demonstrations using continuous-wave pumps, where the reported squeezing is around 2 dB \cite{Mondain19, dutt2015,stokowsky2023, zhao2020} with one experiment reaching 6 dB \cite{Kashiwazaki2020}, and is comparable to the record of squeezing (15 dB) achieved using bulk nonlinear crystals  \cite{vahlbruch2016detection}.

\begin{figure}[!htb]
	\begin{center}
		\includegraphics[width=0.9\columnwidth]{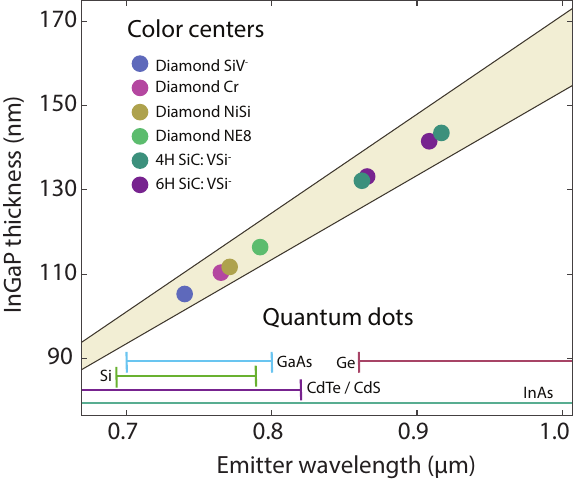}
		\caption{Quantum wavelength conversion to 1550 nm through DFG for solid-state quantum emitters. Shaded area indicates the thickness of InGaP that realizes the phase-matching condition. The emission range of quantum dots are represented using line segments at the bottom. Defect color centers, diamond SiV: \cite{Aharonovich2014}, diamond Cr: \cite{Aharonovich2014}, diamond NiSi: \cite{Steinmetz2011, Rabeau2005}, diamond NE8: \cite{Gaebel2004}, V$^-_{\mathrm{Si}}$ in 4H SiC \cite{Sorman2000, Son2020} and 6H SiC: \cite{Sorman2000}.  Quantum dots, InAs: \cite{Franke2016}, GaAs: \cite{Heyn2009, Zhai2020}, CdTe/CdS \cite{Deng2010}, Si: \cite{Jurbergs2006}, Ge: \cite{Ruddy2010}. }
		\label{fig5:QE}
	\end{center}
\end{figure}

Finally, we consider quantum wavelength conversion using the InGaP nonlinear waveguide. To unify quantum platforms operating at different wavelengths and establish quantum networks utilizing optical fibers, quantum wavelength conversion of photons emitted by matter qubits to telecommunication C band photons is desirable. InGaP has a cutoff wavelength of 650 nm and low optical absorption below the bandgap, which makes it suitable for quantum wavelength conversion involving a variety of quantum emitters. We consider quantum wavelength conversion from quantum emitter wavelengths in the $670-1010$ nm range to 1550 nm via different-frequency generation (DFG) with a pump wavelength longer than the emitter wavelength. We design InGaP-on-insulator (SiO$_2$) waveguides satisfying the phase-matching condition for such DFG processes. To this end, we limit the waveguide width to be in the range between 1 $\mathrm{\mu m}$ and 5 $\mathrm{\mu m}$, and identify the proper thickness of the InGaP such that the phase-matching condition can be realized for a given emitter wavelength. The shaded region in Fig. \ref{fig5:QE} shows the range of desired InGaP thickness depending on the emitter wavelength. Several solid-state defect centers and quantum dots emit in the wavelength range of $670-1010$ nm considered here.  For example, the $\mathrm{SiV^{-}}$ center in diamond has a zero phonon line (ZPL) at 740 nm \cite{Feng1993} and the two types of $\mathrm{V_{Si}^{-}}$ center in 4H-SiC have ZPL at $862$ nm and $917$ nm \cite{Sorman2000, Son2020}. On the other hand, quantum dots can emit into a wide wavelength band depending on their composition and sizes, e.g., InAs ($700-1200$ nm) \cite{Franke2016}, GaAs ($700-800$ nm) \cite{Heyn2009, Zhai2020}, Si ($693-789$ nm) \cite{Jurbergs2006}, CdTe/CdS ($480-820$ nm) \cite{Deng2010} and Ge ($860-1230$ nm) \cite{Ruddy2010}. The quantum efficiency of wavelength conversion from the photon of wavelength $\lambda_1$ to the photon of target wavelength $\lambda_2$ is given by 
\be
F_{\lambda_2}(L)/F_{\lambda_1}(0)=\sin^2(\chi\sqrt{P_p}L/\sqrt{2}),
\ee
where $F_{\lambda}(x)$ is the photon number flux at position $x$ and we have ignored waveguide loss.  Thus the waveguide length to achieve unit-efficiency conversion is 
\be
L_\pi=\pi/(\chi\sqrt{2P_p}).
\ee 
Assuming the nonlinear gain of the waveguide for DFG is similar to that of the waveguide for SHG, we estimate $L_\pi=3.1(9.8)$ mm for pump power $P_p=10(1)$ mW. Such pump powers for high-efficiency quantum wavelength conversion are significantly lower than the state of the art and will reduce parasitic noises that hamper quantum wavelength conversion.    

\section{CONCLUSION}\label{Sec:VI}

In this Perspective, we provided a brief review of the newly developed epitaxial InGaP integrated $\chi^{(2)}$ nonlinear photonics platform and discussed its applications in quantum and nonlinear optics. Leveraging its exceptional nonlinearity and low optical loss, the InGaP platform has achieved highly efficient $\chi^{(2)}$ nonlinear optical effects, surpassing the current state of the art, such as second-harmonic generation and entangled photon pair generation via SPDC. Additionally, the InGaP platform enables novel quantum optical effects at the few- and single-photon levels, which are crucial for quantum information processing and quantum networking. However, addressing certain challenges is necessary to fully realize the potential of the InGaP platform. 
The first challenge is to realize the InGaP-on-insulator platform, which will help mitigate parasitic noises for quantum optics and facilitate high optical powers for applications involving waveguide devices. Recently, the InGaP-on-insulator structure via low-temperature plasma-activated wafer bonding has been reported \cite{thiel2024wafer}. However, the reported optical loss is still worse than the suspended InGaP platform. Further optimization of the wafer bonding process to reduce optical losses is necessary. Another challenge is the thickness nonuniformity of the epitaxial InGaP thin film. The phase-matching condition for  second-order nonlinear optical processes is highly sensitive to the InGaP thickness. This is in general not relevant for microcavities, as the thickness variation occurs at a much larger scale, but will impact waveguides with a considerable length, leading to reduced nonlinear efficiency. Growth condition optimization and the adapted waveguide approach \cite{Chen2024} might be able to overcome this challenge. With these current obstacles addressed, the InGaP $\chi^{(2)}$ nonlinear photonics platform is expected to yield unparalleled performances for a broad range of classical and quantum applications, including optical parametric amplification, squeezed light generation, and quantum wavelength conversion, to name a few.\\

\vspace{2mm}

\noindent\textbf{Data availability}\\ 
All data used in this study are available from the corresponding authors upon reasonable request.

\noindent\textbf{Acknowledgements}\\ 
This work was supported by US National Science Foundation (Grant No. ECCS-2223192), NSF Quantum Leap Challenge Institute QLCI-HQAN (Grant No. 2016136), and U.S. Department of Energy Office of Science National Quantum Information Science Research Centers.

\noindent\textbf{Conflict of interest}\\ 
The authors declare no competing interests.

\vspace{10mm}

	

%

\end{document}